\documentclass[%
reprint,
superscriptaddress,
amsmath,amssymb,
aps,
prapplied,
floats,
]{revtex4-1}  

\usepackage{graphicx}
\usepackage{textcomp}
\usepackage{xcolor}
\usepackage{import}
\usepackage{dcolumn}
\usepackage{bm}

\begin{document}
	
	\preprint{APS/123-QED}

\title{Mutual control of stochastic switching for two electrically coupled superparamagnetic tunnel junctions}

             

\author{Philippe Talatchian}
\email{philippe.talatchian@cea.fr}
\affiliation{Physical Measurement Laboratory, National Institute of Standards and Technology, Gaithersburg, MD, USA}
\affiliation{Institute for Research in Electronics and Applied Physics, University of Maryland, College Park, MD, USA}
\affiliation{Univ. Grenoble Alpes, CEA, CNRS, Grenoble INP, SPINTEC, 38000 Grenoble, France}

\author{Matthew W. Daniels}
\affiliation{Physical Measurement Laboratory, National Institute of Standards and Technology, Gaithersburg, MD, USA}

\author{Advait Madhavan}
\affiliation{Physical Measurement Laboratory, National Institute of Standards and Technology, Gaithersburg, MD, USA}
\affiliation{Institute for Research in Electronics and Applied Physics, University of Maryland, College Park, MD, USA}

\author{Matthew R. Pufall}
\affiliation{Physical Measurement Laboratory, National Institute of Standards and Technology, Boulder, CO, USA}

\author{Emilie Ju\'e}
\affiliation{Associate of the National Institute of Standards and Technology, Boulder, Colorado 80305, USA}
\affiliation{Department of Physics, University of Colorado, Boulder, Colorado 80309, USA}


\author{William H. Rippard}
\affiliation{Physical Measurement Laboratory, National Institute of Standards and Technology, Boulder, CO, USA}

\author{Jabez J. McClelland}
\affiliation{Physical Measurement Laboratory, National Institute of Standards and Technology, Gaithersburg, MD, USA}

\author{Mark D. Stiles}
\email{mark.stiles@nist.gov}
\affiliation{Physical Measurement Laboratory, National Institute of Standards and Technology, Gaithersburg, MD, USA}

\begin{abstract}
Superparamagnetic tunnel junctions (SMTJs) are promising sources for the randomness required by some compact and energy-efficient computing schemes. Coupling SMTJs gives rise to collective behavior that could be useful for cognitive computing. We use a simple linear electrical circuit to mutually couple two SMTJs through their stochastic electrical transitions. When one SMTJ makes a thermally induced transition, the voltage across both SMTJs changes, modifying the transition rates of both. This coupling leads to significant correlation between the states of the two devices.  Using fits to a generalized N\'eel-Brown model for the individual thermally bistable magnetic devices, we can accurately reproduce the behavior of the coupled devices with a Markov model. 

\end{abstract}

\maketitle

\section{Introduction}

Magnetic tunnel junctions have become increasingly applied in nonvolatile memory applications, and are promising building-blocks for novel circuit implementations for cognitive computing~\cite{grollier2020neuromorphic}. These nanoscale devices~\cite{jinnai2020scaling} are composed of two ferromagnetic layers separated by a thin insulating layer. The relative orientation of the magnetization of the two magnetic layers forms two stable configurations with either parallel or antiparallel magnetizations. Magnetoresistive effects \cite{sun2008magnetoresistance} lead to two distinct electrical resistances for the two stable configurations, which can be used to encode memory values of \texttt{0} or \texttt{1}. Applying voltages across these nanojunctions results in spin-polarized tunneling currents that apply spin-transfer torques~\cite{berger,slonczewski,ralph2008spin} on the magnetizations. For sufficiently large voltages, the energy barrier between the two configurations is overcome, leading to magnetization switching events and allowing memories to be written. 

In order to obtain the years of retention required by nonvolatile applications, the energy barrier between the two memory states needs to be large, greater than 40 $kT$, where $k$ is Boltzmann's constant and $T=300$~K is room temperature. Reducing the energy barrier decreases the retention time exponentially. For energy barriers smaller than $\approx 12~kT$, the thermal fluctuations at room temperature induce random switching of the magnetization between the two stable configurations on the 10~{\textmu}s to 1~ms time scale. A magnetic junction with its magnetic configuration fluctuating due to a small energy barrier is referred to as a superparamagnetic tunnel junction (SMTJ).

While the switching behavior of SMTJs is inherently random, the average relative dwell time spent in each stable configuration before a stochastic switching event occurs can be tuned deterministically~\cite{rippard2011thermal}. This tunability can be achieved by applying a magnetic field to alter the energy barriers, or by applying a voltage across the junction to induce a spin-transfer torque that favors one of the two configurations. Tunability can also be achieved by passing a current through an adjacent heavy metal~\cite{lee2016emerging} to create a spin-orbit torque~\cite{garello2013symmetry}. 

SMTJ devices share many of the practical advantages of their nonvolatile MRAM counterparts. In particular, they are compatible with complementary metal oxide semiconductor (CMOS) technology and can be fabricated in large numbers at competitive densities~\cite{chung_4gbit_2016}. These properties make SMTJs promising candidates for building compact, low-energy random number generators~\cite{parks2018superparamagnetic,PhysRevApplied.8.054045}, and make them attractive for efficient, unconventional computing schemes like probabilistic~\cite{camsari2019p, borders2019integer} or brain-inspired~\cite{mizrahi2018neural, daniels2020energy, sengupta2016probabilistic, cai2019voltage} computing. 

Parallel to these efforts, spintronic nano-oscillators based on non-stochastic magnetic tunnel junctions are also being investigated as  candidates for neuromorphic computing~\cite{torrejon_neuromorphic_2017, romera2018vowel, zahedinejad2020two}. Since SMTJs can be seen as stochastic bistable ``oscillators,'' many computing schemes developed for these nano-oscillators may in fact be adaptable to SMTJs. \textcolor{black}{Neuromorphic applications like reservoir computing, for instance, have been developed based on stochastic neurons~\cite{verstraeten2005reservoir}}. One potential advantage that SMTJs may have over spintronic nano-oscillators is lower energy consumption~\cite{daniels2020energy}.  Another is that the time scale of the oscillations can be easily tuned over several orders of magnitude to match the time scale needed in real-time applications.
 
An essential property of traditional spintronic nano-oscillators is their ability to modify their frequency and phase naturally and synchronize by receiving the emitted stimuli of other nano-oscillators or external periodic drives. Mimicking computational schemes based on this property with superparamagnetic tunnel junctions requires demonstrating that the latter have similar abilities.  Locatelli \emph{et al.}~\cite{locatelli2014noise} have shown experimentally that a single superparamagnetic tunnel junction can exhibit stochastic synchronization to a periodic external signal. This effect can be enhanced by an optimal electrical noise level~\cite{cheng2013nonlinear,mizrahi2016controlling} corresponding to a stochastic resonance~\cite{gammaitoni1998stochastic,neiman1998stochastic}. The next step is to demonstrate and characterize the coupling between two superparamagnetic tunnel junctions.  

In this work, we use a straightforward electrical circuit to establish an electrical interaction between two SMTJs without complex circuitry, nonlinear elements, or additional computation. We show that this mutual coupling correlates their switching events. 

Section~\ref{sec:schematic} presents the electrical scheme we use to couple the two SMTJs and describes the coupling mechanism. Section~\ref{sec:smtjs} gives the characterization of the uncoupled SMTJs. This preliminary step is necessary for extracting the individual device parameters. Section~\ref{sec:dccoupling} reports measurements of the behavior of coupled SMTJs. We compute the correlation functions between the two SMTJs and demonstrate that their states are correlated. We show that we can predict their behavior with simple models described in the appendices. We discuss the role device and circuit properties play in the coupling in Sec.~\ref{sec:discussion}, and show how the models developed can be used to predict the behavior of larger networks of SMTJs. Appendix~\ref{app:neelbrown} introduces the N\'eel-Brown model~\cite{brown1963thermal} we use to analyze the mean dwell-times and parameterize the behavior of uncoupled devices. Appendix~\ref{app:markov} describes the Markov model that uses the parameters taken from the N\'eel-Brown model fits and computes the correlated behavior of the coupled devices.

 \section{Coupling between superparamagnetic tunnel junctions}
 \label{sec:schematic}
 
The majority of coupling schemes reported for two or more SMTJs require protocols that invoke additional peripheral circuits~\cite{camsari2017stochastic, roy2018perspective, debashis2020correlated} or finely-calibrated external drive stimuli~\cite{mizrahi2016synchronization}. These complexities make such schemes problematic for large, generalized networks. We employ a simpler approach to coupling SMTJs based on combinations of linear circuit elements. Our intention is that this will offer a more robust, less complicated, and more compact foundation for future explorations of large scale SMTJ networks. While expanding the two-SMTJ coupling scheme discussed here to large networks of SMTJs will most likely require more complicated circuitry, we believe that understanding of the minimal coupling needed will help with designing the simplest circuitry possible.

\begin{figure}
	\includegraphics[width=\columnwidth]{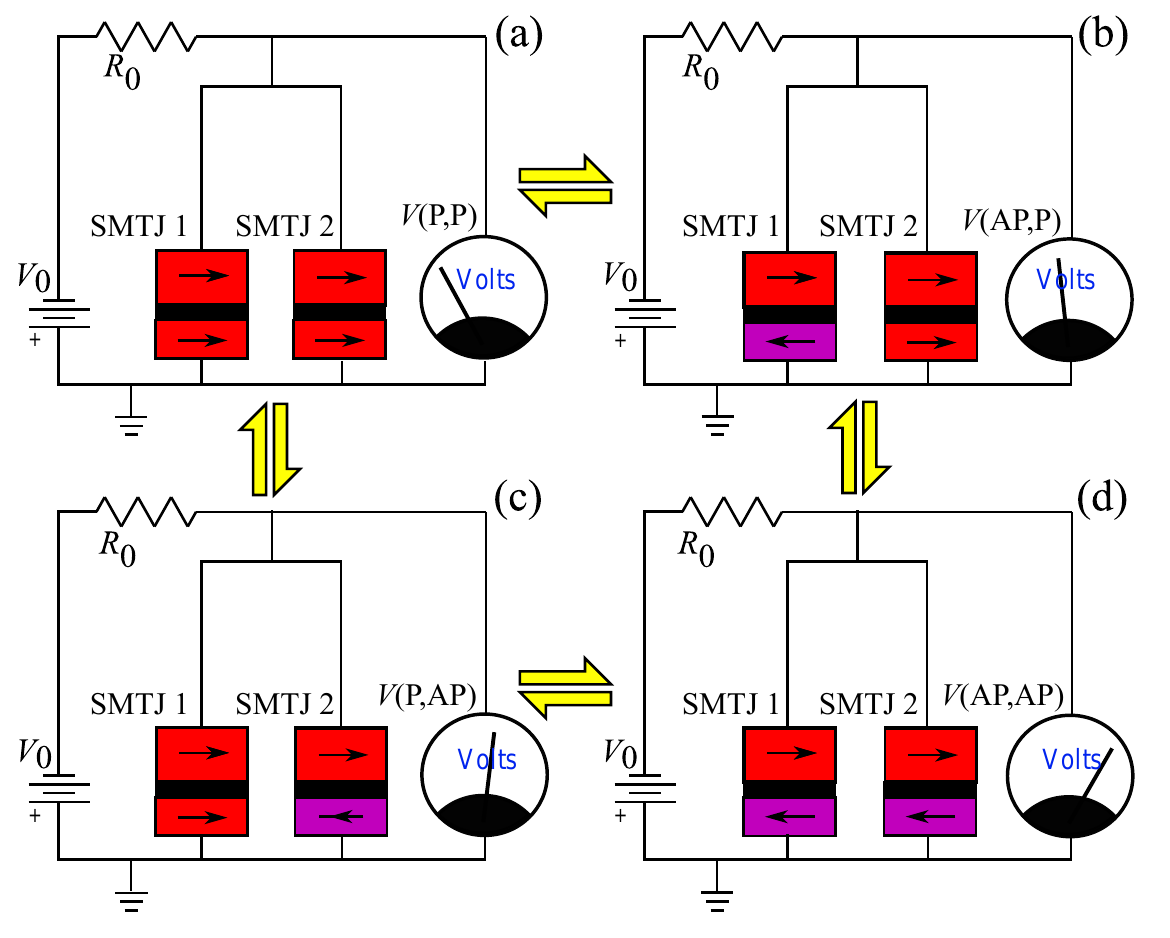}
	\caption{\label{fig:circuit} Schematic of the electrical circuit used to couple two superparamagnetic tunnel junctions and the four metastable configurations they take.  Each panel shows the constant voltage source $V_0$, the series resistor $R_0$, the two SMTJs in parallel with each other, and a voltmeter representing the oscilloscope measurement of the shared voltage across the SMTJs. The yellow arrows connecting the four panels represent thermally driven transitions between the different configurations of the SMTJs. Panel (a) shows both SMTJs in the parallel (P) state, with the free layer (bottom) magnetization parallel to the fixed layer (top) magnetization.  When the magnetization in the free layer of either SMTJ flips, panels (b) and (c), the resistance of the now antiparallel SMTJ (indicated by a purple free layer) increases so the voltage across both SMTJs increases. The voltages in these two cases differ due to differences in the magnetoresistance of the two devices. If both free layers flip, the system goes to the configuration in panel (d), with both SMTJs in the antiparallel (AP) state and the largest magnitude voltage across both SMTJs. 
	}
\end{figure}

Figure~\ref{fig:circuit} shows the electrical circuit we use to induce interactions between two SMTJs. We connect the two SMTJs in parallel and bias them through a series resistor $R_0$ with a constant voltage source $V_0$. This circuit results in four distinct voltages applied across the SMTJs, corresponding to the four different magnetization configurations in the system: (P,~P), (AP,~P), (P,~AP), and (AP,~AP), where P denotes the parallel, low resistance state of the SMTJ and AP denotes the antiparallel, high resistance state. These states are illustrated respectively in panels (a), (b), (c), and (d) of Fig.~\ref{fig:circuit}. The four voltages are different because in each case the effective resistance of the two parallel SMTJs is different, resulting in a greater or lesser portion of $V_0$ being dropped across the series resistor $R_0$. Though the configurations (P,~AP) and (AP,~P) would have the same voltage if the two SMTJs were identical, in practice the P and AP resistances of our two SMTJs differ somewhat, allowing the voltage drops to be distinguished between the two mixed states.  Because the voltage is a property of the joint state of the devices, a switch in either device's configuration induces a change in voltage across the other device. This joint voltage dependence effectively couples the behavior of each SMTJ to the other.            

The SMTJs can be placed in the circuit with their free layer attached to either the positive or negative terminal of the dc voltage source.  In the positive case, electrons flow from the fixed layer to the free layer and a higher voltage tends to stabilize the parallel magnetization configuration. Conversely, with the polarity reversed, the antiparallel configuration is stabilized by higher voltages.  We choose the positive case so that when an SMTJ switches from parallel to antiparallel, the increase in voltage magnitude destabilizes the antiparallel state and vice versa. This choice induces a positive correlation between the states of the SMTJs as we show in Sec.~\ref{sec:dccoupling}.

\textcolor{black}{To understand the coupling, consider SMTJ1 in the parallel configuration and consider how the relative stability of that state depends on the configuration of SMTJ2. If SMTJ2 is in the parallel configuration, the voltage across both SMTJs is lower than it would be were SMTJ2 in the antiparallel configuration. Because of the lower voltage, SMTJ1 is relatively more stable in the parallel configuration, increasing the lifetime of \textcolor{black}{the} (P,~P) configuration relative to that of the (P,~AP). Similarly, if SMTJ1 is in the antiparallel configuration, the changes in the voltage for the different configurations of SMTJ2 increase the lifetime of the (AP,~AP) configuration relative to that of the (AP,~P) because higher voltages favor the AP state. Repeating the analysis from the point of view of SMTJ2 gives the same result -- for the choice of polarity made here, the changes in the voltage as the resistances change with configuration increase the lifetimes of the (P,~P) and (AP,~AP) configurations relative to the dissimilar configurations. The changes in these lifetimes is the origin of the coupling.}

To read out the states of the two coupled SMTJs in the electrical circuit, we monitor the common voltage across the two SMTJ branches shown in Fig.~\ref{fig:circuit} on an oscilloscope. From the measured voltage, we can determine the configuration of both SMTJs, the time they spend in those configurations, and which SMTJ makes a transition to another configuration.

The statistics of the transitions between the configurations can be described by a Markov model, described in Appendix~\ref{app:markov}, using the rates computed in Appendix~\ref{app:neelbrown}. We treat each pair of configurations of the SMTJs as a state and compute the two transition rates out of each state.  We find that this model provides a good explanation of the measurements subject to the accuracy of the N\'eel-Brown model fit for the rates.

\section{Superparamagnetic tunnel junctions}
\label{sec:smtjs}
 
We study magnetic tunnel junctions with the following composition: Si substrate / SiO$_2$ / Ta(3) / Cu(10) / Ta(3) / Cu(3) / IrMn(10) / CoFe(3) / Ru(0.8) / CoFeB(3) / CoFe(0.2) / MgO(1.08) / CoFe(0.2) / CoFeB(1.8) / Ta(4) / Ru(10) / Ta(4). Numbers in parentheses represent thicknesses in nanometers. Here  IrMn(10) / CoFe(3) / Ru(0.8) / CoFeB(3) is a uniformly in-plane magnetized synthetic ferrimagnet that plays the role of polarizer. CoFe(0.2) / CoFeB(1.8) is the free-layer. The devices are elliptical with nominal dimensions of 60~nm $\times$ 72~nm. Their tunneling magnetoresistance (TMR) value is close to \textcolor{black}{55~\%} at room temperature and they have a resistance-area product of 4.75~$\Omega$~{\textmu}$\text{m}^2$ \footnote{The sample was prepared by sputter deposition in a system with base pressure of 1.33$\times 10^{-7}$~Pa (10$^{-9}$~Torr). The Mg barrier is oxidized in two steps. 0.6~nm Mg is deposited, then oxidized in the chamber load lock at an O$_2$ pressure of $\approx9.3\times10^{-2}$ Pa ($7\times 10^{-4}$~Torr) for 100~s. The sample is returned to the chamber, and 0.45~nm Mg deposited, then returned to the load lock and oxidized for 420~s at an O$_2$ pressure of $\approx0.27$ Pa ($2.8\times10^{-3}$~Torr). The sample is then returned to the chamber and 0.3~nm Mg deposited, followed by the rest of the listed stack. Post deposition, the sample is annealed in vacuum at 350~$^\circ$C for 1~h in a 1~T in-plane field to set the reference layer pin direction and recrystallize the barrier. The pinning field direction is nominally along the long axis of the elliptical devices.}.
 
\begin{figure}
	\includegraphics[width=\columnwidth]{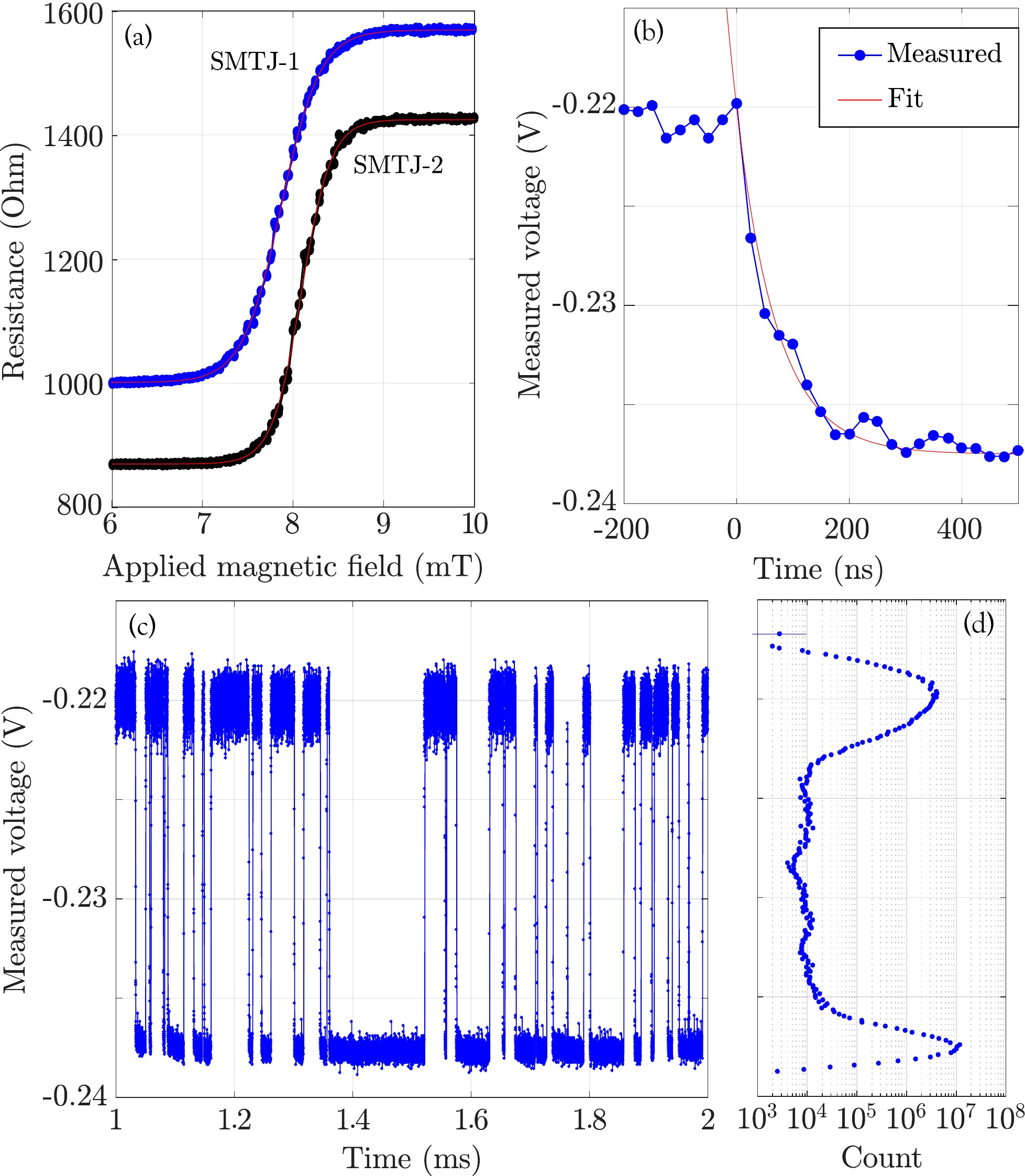}
	\caption{\label{fig:single} (a) Experimental evolution of the dc resistance of the studied SMTJs versus the applied in-plane magnetic field, swept from 6~mT to 10~mT and back. Those characteristics were obtained for a constant applied dc voltage $V_0=-0.1$ V and a series resistor of $R_0=473~\Omega$. (b) Experimental time trace of the voltage of a single SMTJ $\mu_0H= 6.9$~mT, $V_0=-0.45$~V, in parallel with a static resistor of $R_{eq}=1085~\Omega$. This panel covers a single transition from the antiparallel to the parallel state illustrating the effects of $RC$ time constants on the transitions.  Fit to an exponential gives a time constant $\tau_{RC}$ for the measurement of (71$\pm$14)~ns. (c) Longer time trace showing approximately 50 transitions over 1~ms. \textcolor{black}{The different noise levels around the two voltage states reflect that the measurement is being taken in an asymmetric configuration; the lower voltage state is relatively stabilized by the applied voltage, and consequently undergoes weaker thermal fluctuations.} (d) Histogram of measured voltages over 2~s containing roughly $10^5$ transitions. Note logarithmic $x$-axis; $y$-axis is shared with (c). }
\end{figure}

The measurement setup for uncoupled SMTJs consists of an individual SMTJ in series with a resistor with constant dc voltage applied to the circuit.  This allows the voltage to vary across the SMTJ as its resistance changes. \textcolor{black}{The series resistor is chosen to maximize the voltage swing between states, as best as possible given the two different devices. Note that this is not necessarily the optimal choice to maximize correlation once we couple the devices; we investigate this question in Sec.~\ref{sec:discussion}.} A constant in-plane magnetic field $H$ is applied to the device along its easy axis. For both SMTJs measured here, with an applied dc voltage $V_0\approx 0$, applied magnetic fields near $\mu_0H\approx 8$~mT make both devices thermally unstable in both the antiparallel (AP) and parallel (P) configurations.  This behavior is apparent through the non-hysteretic resistance curves of the devices shown in Fig.~\ref{fig:single}(a), in which the magnetic field was ramped slowly enough to measure the time-averaged resistance at each field value. 

In order to evaluate the mean dwell times, $10^4$ to $10^5$ transitions were recorded and analyzed for each field value and each voltage. Fig.~\ref{fig:single}(b) shows a typical time trace around a state transition from parallel to antiparallel. The effective $RC$ time constant of the oscilloscope manifests as the exponential decay in Fig.~\ref{fig:single}(b). Dwell times shorter than this $RC$ decay can affect the apparent mean dwell times of the devices~\footnote{Dwell times shorter than the $RC$ decay would affect our results in two ways. First, the associated transitions could get hidden from our histogrammatic analysis and not properly counted. Second, transitions during this period actually follow a parametrically different Poisson process from the two we nominally expect to see, because the instantaneous voltage across the device corresponds to neither the P nor AP steady states.}. Fig.~\ref{fig:single}(c) shows a typical time trace at a larger time scale. Dwell times greater than $\approx 1\;${\textmu}s, such as those visible in Fig.~\ref{fig:single}(c), are unaffected by the $RC$ dynamics.

We analyze the time traces to determine the properties of the SMTJs. Simply binning the data in the time traces gives the probabilities to be in each state, as in Fig.~\ref{fig:single}(d). These probabilities do not depend on determining the length of time spent in each state. Determining the mean dwell time for each state requires that we determine every intertransition interval $i$, and extract the corresponding dwell times $\Delta t_i^{\pm}$ for the anitparallel ($+$) or parallel ($-$) states. The cumulative probability density function of these extracted dwell times can be fit by an exponential distribution. The single fitting parameter of the exponential distribution is the mean dwell time $\tau^{\pm}=\langle\Delta t_i^{\pm}\rangle_i$, where $\langle \cdots \rangle_i$ indicates averaging over all intervals.

By varying the applied in-plane magnetic field $H$ and the applied dc voltage $V_0$, we observe variation in the dwell times. For each pair $(V_0,H)$, the parallel $\tau^{-}$ and antiparallel $\tau^{+}$ dwell times are extracted from the corresponding voltage-time trace. The evolution of the dwell times as a function of the applied in-plane magnetic field is reported in Appendix~\ref{app:neelbrown}. The magnitude of $H$ determines whether the parallel $\tau^{-}$ or antiparallel $\tau^{+}$ mean dwell time dominates. In Appendix~\ref{app:neelbrown}, we fit the experimental values of the mean dwell times to the N\'eel-Brown model~\cite{brown1963thermal}. \textcolor{black}{We find that the usual N\'eel-Brown model is unable to capture the behavior of our devices when both field and voltage dependence are considered. We extend the model slightly by allowing the characteristic magnetic fields in the model to have a state dependence, prescribing different values depending on whether the device is in the P or AP state. This extension gives our fit good agreement with the data, but we caution that the fit---while physically motivated---may be highly degenerate and does not necessarily prescribe physically meaningful values to the model parameters. We discuss this further at the end of Appendix~\ref{app:neelbrown}.}

\textcolor{black}{Finally, note that the TMR observed in Fig.~\ref{fig:single}(a) appears to be inconsistent with the results in Fig.~\ref{fig:single}(c). One reason for this is the voltage dependence of the magnetoresistance; the higher voltage applied to the devices in Fig.~\ref{fig:single}(c) results in a lower TMR. It is also possible that properties of the devices or their wire bonds changed slightly between characterization [Fig.~\ref{fig:single}(a)] and experiment [Fig.~\ref{fig:single}(c)]. In any case, we find that the two voltage states such as those in Fig.~\ref{fig:single}(c) are stationary (for each chosen source voltage) across the voltage-time trace experiments and show no evidence of a third state. In fitting the N\'eel-Brown model, we fit directly to these voltage states, bypassing any functional dependence on TMR.}

\section{Interaction of two superparamagnetic tunnel junctions through electrical dc coupling}
\label{sec:dccoupling}

In the previous section, we described the switching statistics of single SMTJs in series with a static resistor. In the present section, we place another SMTJ in parallel with the first, as in Fig.~\ref{fig:circuit}. Our goal is to extract the switching data from this experimental setup and use it to determine whether there is statistically meaningful coupling between the devices.

\begin{figure}
	\includegraphics[width=\columnwidth]{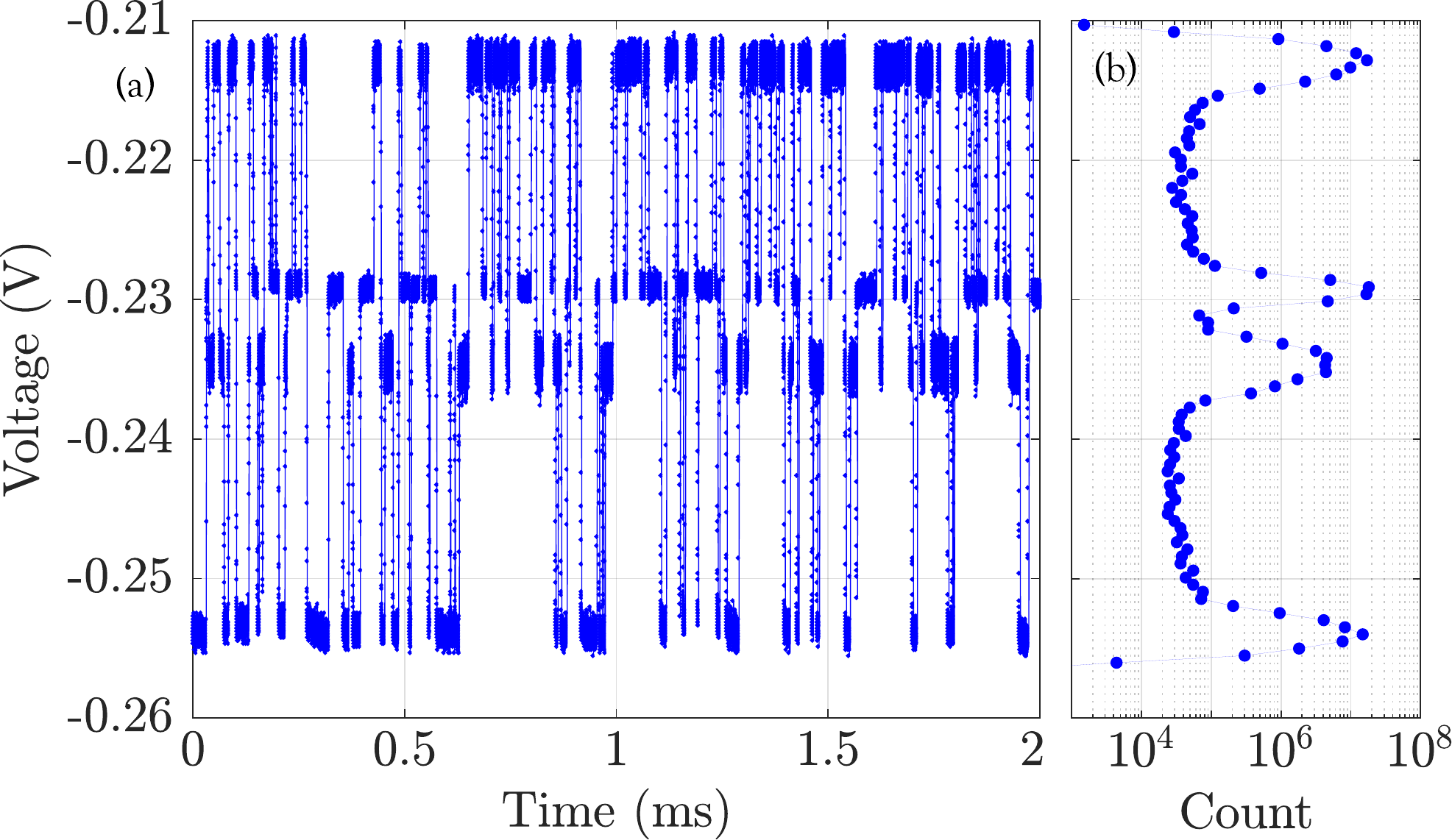}
	\caption{\label{fig:coupledTimeTrace} (a) Time trace of the voltage of two coupled SMTJs, for $\mu_0H=6.9\;\text{mT}$, $V_0=-0.45\;\text{V}$ and $R_0=473~\Omega$. (b) Histogram of measured voltages over 20~s. Note the logarithmic scale on the horizontal axis.  }
\end{figure}

Figure~\ref{fig:coupledTimeTrace}(a) shows a typical time-voltage trace of the circuit from Fig.~\ref{fig:circuit} under relevant experimental parameters. We again use a histogram method to extract the multiple states; there are now four such states corresponding to the four panels of Fig.~\ref{fig:circuit}. Figure~\ref{fig:coupledTimeTrace}(b) shows the histogram of a voltage-time trace, revealing four distinct peaks. The integrated area under each peak gives the relative probability of the system being found in the corresponding state. 

\begin{figure}
	\includegraphics[width=\columnwidth]{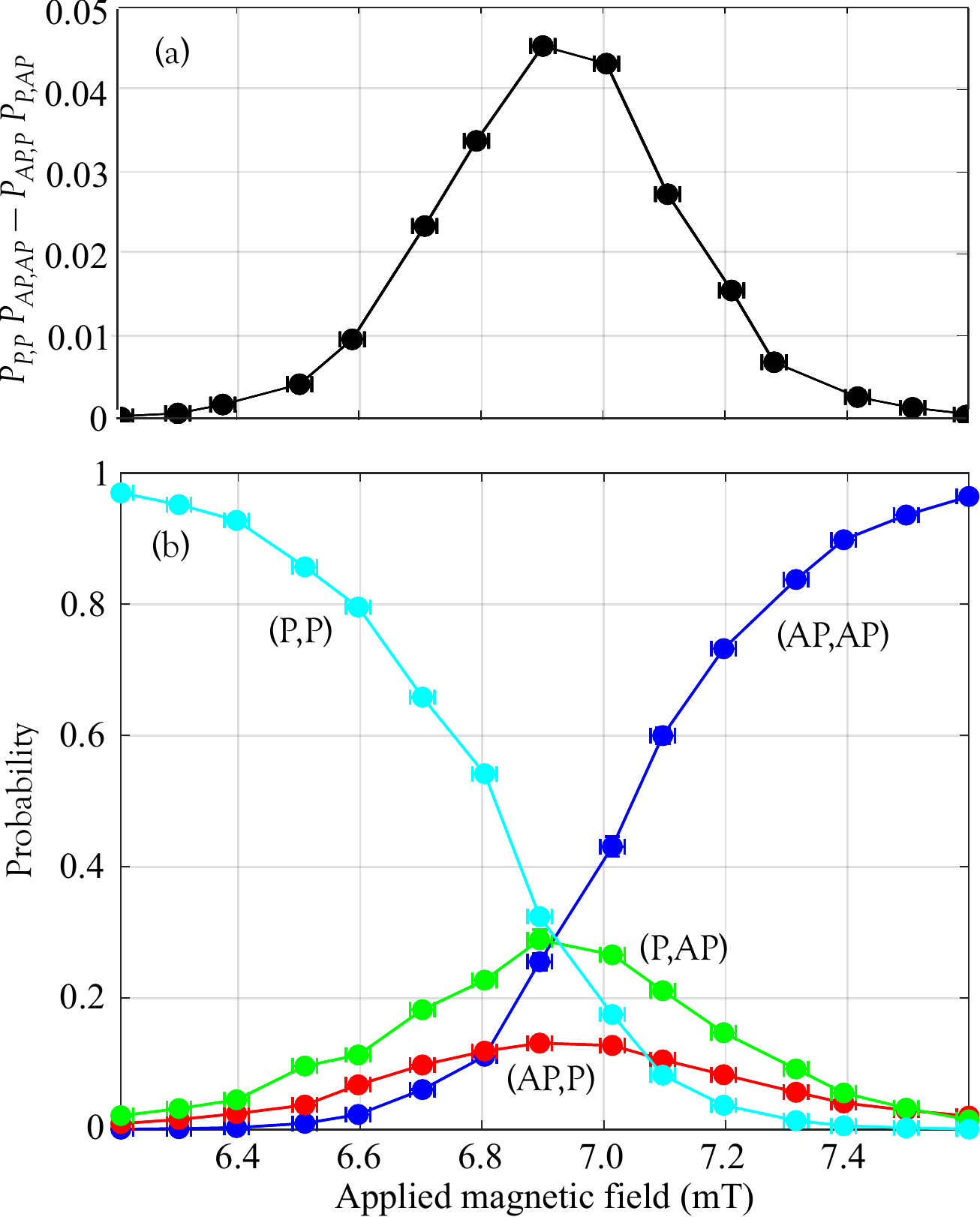}
	\caption{\label{fig:occupancyProbability} (b) Probabilities of occupancy for the different joint configurations of two coupled SMTJs, with $V_0=-0.45\;\text{V}$. (a) Determinant of the probabilities in (b) as in Eq.~\eqref{eq:p-matrix}. Error bars indicate single standard deviation uncertainties.}
\end{figure}

By computing the probabilities over a range of magnetic fields, we identify the superparamagnetic regime between two deterministic limits, one limit with field large enough that both SMTJs are pinned in the antiparallel state, and the other with a low enough field that they are both pinned in the \textcolor{black}{parallel} state. The probabilities are plotted for this regime in Fig.~\ref{fig:occupancyProbability}. At low magnetic field, the parallel states of both devices are stable; at high magnetic field, on the other hand, the antiparallel states are stabilized. The superparamagnetic regime exists between these two limits, where the mixed states (P,AP) and (AP,P) acquire nonzero lifetimes. 
The qualitative behavior of Fig.~\ref{fig:occupancyProbability} could be explained by the uncoupled-device physics we discussed in Sec.~\ref{sec:smtjs}. Quantitatively, the probabilities show significant correlation between the switching events in the two devices.

\begin{figure}
	\def\svgwidth{\columnwidth}
    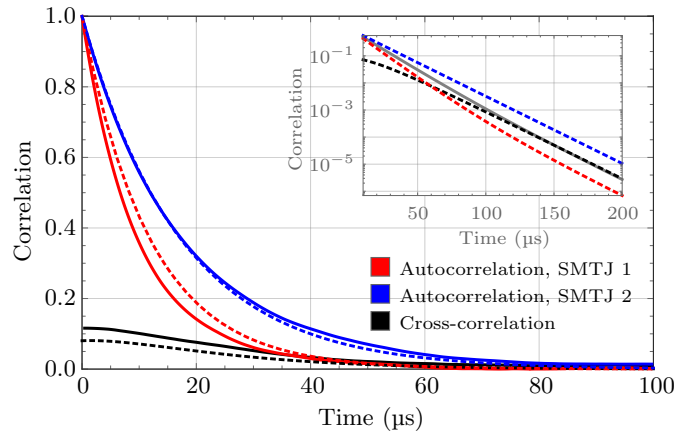
	\caption{\label{fig:correlationFits} Auto- and cross-correlation for the states of two coupled SMTJs.  Solid lines are derived from experimental data at $\mu_0H= 7.6$~mT, $V_0=-0.3$~V. Dashed lines are derived from the N\'eel-Brown model fit with the parameters of Table~\ref{tab:table1} (Appendix~\ref{app:neelbrown}) at the same field and voltage, and a Markov model for the coupling (Appendix~\ref{app:markov}). In the log scale inset, the gray line is the geometric mean of the model-predicted coupled autocorrelations, which serves as an asymptotic upper bound for the cross-correlation.} 
\end{figure}

To understand the coupling between the SMTJs we evaluate the correlation functions between the encoded times series data of SMTJs defined by taking the value $-1$ when the SMTJ is in the parallel state and $1$ when it is in the antiparallel state. We then compute the normalized Pearson correlation function, which for two arbitrary functions of time, $x(t)$ and $y(t)$, is given by
\begin{align}
    C_{x,y}(t)&=  \frac{\langle [ x(t')-\langle x \rangle ] [ y(t'+t)-\langle y \rangle ] \rangle}{\sqrt{\langle [ x(t')-\langle x \rangle]^2 \rangle \langle [ y(t')-\langle y \rangle]^2 \rangle}},
    \label{eq:correlation}
\end{align}
where $\langle \cdots \rangle$ indicates average over $t'$. When computing an autocorrelation function, $x(t)$ and $y(t)$ are the same encoded times series shifted in time and when computing a cross-correlation function they are from different SMTJs. Fig.~\ref{fig:correlationFits} shows, with the solid curves, the autocorrelation and cross-correlation functions computed using time-series state data extracted from the experiment for a particular magnetic field and voltage. The presence of a nonzero cross-correlation function indicates correlation between the random variables corresponding to each SMTJ's state.
The Markov model described in Appendix~\ref{app:markov} shows that these correlation functions are sums of three exponential functions of time. 
The model's prediction of correlation based on the uncoupled fits is given by the dashed curves in Fig.~\ref{fig:correlationFits}.

To test for coupling, we are particularly interested in the $t=0$ value of the cross-correlation function, which depends only on probabilities like those in Fig.~\ref{fig:occupancyProbability}. Those probabilities can be easily extracted from a histogram analysis and do not require extensive identification of individual transitions. Therefore, our method to compute the $\tau=0$ cross-correlation is particularly easy and suitable to quantify coupling between SMTJs by experimentalists. The $t=0$ form of Eq.~\eqref{eq:correlation} reduces to Eq.~\eqref{eq:correlation-formula}, the numerator of which is equal to four times the determinant of 
\begin{equation}
    \mathcal P = \left(\begin{matrix}
    P_\text{P,P} & P_\text{P,AP}\\
    P_\text{AP,P} & P_\text{AP,AP}
    \end{matrix}\right).\label{eq:p-matrix}
\end{equation}
When the two SMTJs are statistically independent, each of these four joint probabilities factor so that $\det \mathcal P = 0$. A nonvanishing determinant indicates mutual dependence of the SMTJ states in a way that can be deduced from the data in Fig.~\ref{fig:occupancyProbability}(b) and shown in Fig.~\ref{fig:occupancyProbability}(a). For example, at $\mu_0H=6.84$~mT, the (AP,P) state is significantly less probable than the other three state so that the skew diagonal term in $\det\mathcal P$ is less than the diagonal term and a positive correlation results. In general, for this voltage polarity, the coupling reduces the mean dwell time for each of the mixed configurations relative to what it would be if there were no coupling. If we were to do a similar measurement with the voltage source reversed, the coupling would increase the mean dwell time for each of the mixed configurations, leading to a negative value for the correlation function.



\begin{figure}
	\def\svgwidth{\columnwidth}
    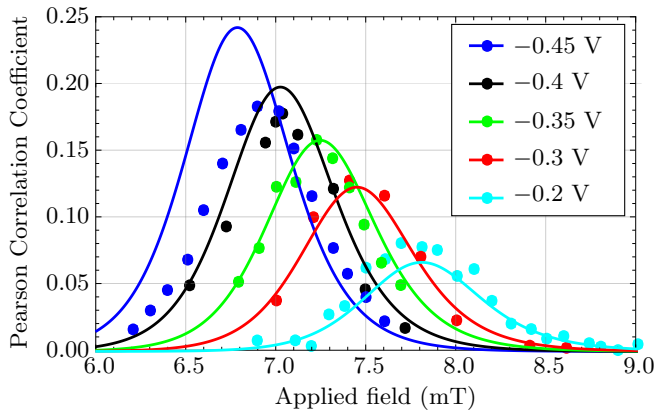
	\caption{\label{fig:couplingTrends} The correlation coefficient (equivalently the $t=0$ cross-correlation) between the states of the coupled SMTJs at a series of source voltages $V_0$. Filled circles show the experimental data; curves are the model predictions based on the Markov model described in Appendix~\ref{app:markov} using the N\'eel-Brown fits to isolated SMTJs described in Appendix~\ref{app:neelbrown}. The single standard deviation uncertainties for the experimental points are smaller than the symbols. They are a result of the finite number of intervals measured in each state.}
\end{figure}
Figure~\ref{fig:couplingTrends} shows the $t=0$ cross-correlation as a function of field for various voltages. Comparing the $V=-0.45$~V data in Fig.~\ref{fig:couplingTrends} with the data in Fig.~\ref{fig:occupancyProbability}, we infer that maximal coupling happens in the middle of the superparamagnetic regime, namely near where the mixed states (AP,~P) and (P,~AP) have their highest probabilities. The coupled system is most sensitive when both SMTJs have balanced occupancies because the mixed configurations become the most probable. Figure~\ref{fig:single}(a) also shows that the two SMTJs used in this experiment have these balance points shifted with respect to each other. \textcolor{black}{This shift reduces the coupling between the devices.} 

\textcolor{black}{Simulations show that the more similar the properties of the two SMTJs are, the stronger the coupling between them. Specifically, if the derivatives of the results in Fig.~\ref{fig:single} are viewed as susceptibilities, the coupling to each devices is maximum at the peak in the susceptibility, so that the mutual coupling is largest when the peaks in the susceptibility align with each other. Further, similar susceptibility-like curves can be measured as a function of current for fixed fields. The coupling will be strongest when these curves align as well. Thus, it is important to keep device properties that affect these susceptibilities as similar as possible to maximize this coupling, but constraints on other properties, like the exchange stiffness in the limit of single domain switching, are less important. } 

The balance point between the two configurations shifts in magnetic field as the voltage magnitude increases. This trend is illustrated in Fig.~\ref{fig:couplingTrends}.
As the voltage magnitude increases, the stability of the AP state increases, reducing the magnetic field needed to balance the configurations. In addition to the peak shifting to lower magnetic fields, it also increases in magnitude because the greater source voltage $V_0$ results in greater swings in voltage drop across the SMTJs as their configurations fluctuate. Note that sustained increase depends on the equal-probability magnetic fields for the two SMTJs changing in the same way as $V_0$ increases.

In Appendix~\ref{app:neelbrown}, we fit the behavior of each SMTJ to a N\'{e}el-Brown model to capture the mean dwell times of the individual devices as a function of magnetic field and voltage. For a given voltage and magnetic field, this model provides the mean transition rates that enter into the Markov model for the coupled states described in Appendix~\ref{app:markov}. Using this model, we compute the correlation function corresponding to the measurements shown in Fig.~\ref{fig:couplingTrends} as symbols and plot the computed values as curves. For low voltages, the agreement is quite good, but it degrades for higher voltages, where the voltages across the SMTJs exceed those used for the fits that feed into the Markov model. 



\section{Discussion}
\label{sec:discussion}


Among the device properties that determine how strongly the behaviors of two SMTJs correlate with each other are the TMR and how similar the two devices are. These determine how much a change in the state of one of the SMTJs affects both of them by determining the resulting voltage swings. The more similar the devices, the more their susceptibility overlaps, making it easier for correlations to develop. Each SMTJ is most sensitive to the other SMTJ's state at magnetic field and mean voltage combinations for which the probability of being in either state is close to 50~\%. This is illustrated by comparing the curves in Fig~\ref{fig:occupancyProbability} with the results for $V_0=-0.45$~V in Fig.~\ref{fig:couplingTrends}. While the two SMTJs used here differ somewhat in their properties, as seen in the shift in the 50/50 points in Fig.~\ref{fig:single}, they are similar enough \textcolor{black}{so as not to} significantly reduce the maximum observed correlation. 

\begin{figure}
	\def\svgwidth{\columnwidth}
    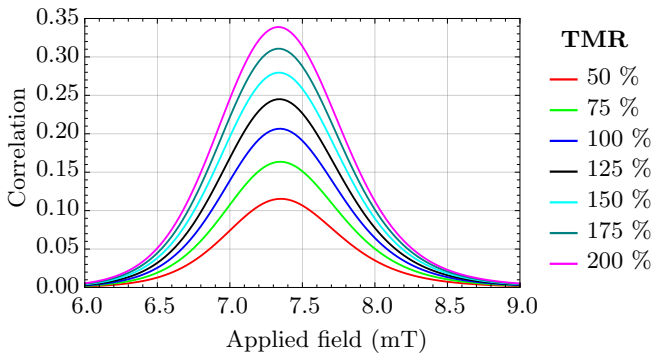
	\caption{\label{fig:tmrTrends} Model predictions of the correlation coefficient (equivalently the $t=0$ cross-correlation) between the states of coupled SMTJs at a series of tunneling magnetoresistances demonstrating that greater TMR gives stronger correlation. The SMTJs are identical and based on the experimental SMTJ 1.}
\end{figure}

Figure~\ref{fig:tmrTrends} show that the TMR substantially affects the coupling. We simulate two identical SMTJs, both the same as SMTJ-1 with the exception that they have a series of TMR values. The model was evaluated at $V_0=-0.3$~V, with $R_P$ and $R_{AP}$ changing as a function of TMR so that the mean equivalent resistance of the four Markov states is held constant at each TMR~\footnote{Under this constraint, $R_P$ can be expressed as a function of the TMR $T$, with $R_P = [(2+T)/(8+8T+T^2)] \times 4159\;\Omega$.}.  Between TMR values of 50~\% to 200~\%, the maximum correlation increases by about a factor of three. As the TMR varies, the voltage swings due to state changes increase, increasing the coupling. Along similar lines, increasing the spin-torque efficiency increases the coupling by increasing the sensitivity to voltage swings.

\begin{figure}
	\def\svgwidth{\columnwidth}
    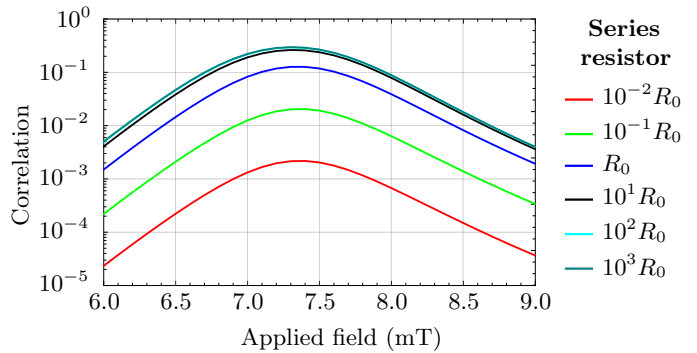
	\caption{\label{fig:resistorTrends} Model predictions of the correlation coefficient (equivalently the $t=0$ cross-correlation) between the states of coupled SMTJs over a sequence of series resistor values (as multiples of $R_0 = 473~\Omega$) values wherein the mean voltage drop across the four Markov states is fixed. The data demonstrates that changing from a voltage to a current source gives stronger correlation. The model was evaluated with two identical SMTJs based on the experimental SMTJ 1. The cyan curve ($10^2 R_0$) is covered almost entirely by the teal curve ($10^3 R_0$), demonstrating saturation behavior of the correlation at high series resistance.}
\end{figure}

Two circuit properties that determine how strong the coupling is are the applied voltage and the series resistance. Varying these properties together switches between voltage biasing and current biasing the SMTJs. These trends are illustrated in Fig.~\ref{fig:resistorTrends}, where $R_0$ is systematically varied with $V_0$ adjusted to keep the mean voltage across the SMTJs fixed. As $R_0$ goes to zero, which is the voltage bias limit, the correlation goes to zero because the voltage across the SMTJs does not change as the configurations change. As $R_0$ increases, the maximum correlation approaches around $0.3$. This $R_0\rightarrow \infty$ limit is the regime where the circuit is driven by a current source, and gives maximal coupling between the devices, at the expense of decreasing the bandwidth of the circuit as the $RC$ time constant diverges with $R_0$. These results indicate that there is a trade-off between the the size of the signal and the speed of the circuit going between the current-biased limit and the voltage biased limit. 

The SMTJs we work with here require external magnetic fields on the order of 6~mT to 9~mT to be in the superparamagnetic regime for the voltages used. The necessity of such fields would limit the use of this type of SMTJ in practical applications. However, it is possible to tune these offset fields by engineering the materials stack of the SMTJs \cite{devolder2019offset}. Recent developments in spin-orbit-torque-switched MRAM \cite{garello2019manufacturable} show that it is also possible to engineer a magnetic element above the MTJ. Designing the SMTJs with magnetic elements such that no additional applied magnetic field would be required to center the SMTJ's 50/50 point around the operating voltage would give the largest effect for the lowest energy expenditure.

Another direction to explore for exploiting the SMTJ coupling examined in this paper involves scaling up to chains or grids of devices. The behavior of such networks can be explored quasi-analytically using the Markov model described in Appendix~\ref{app:markov} or simulations using the N\'{e}el-Brown fits described in Appendix~\ref{app:neelbrown}. In Fig.~\ref{fig:5chain}, we compare the behavior predicted from these two approaches for a chain of five identical SMTJs electrically coupled in parallel analogously to the coupling of two in Fig.~\ref{fig:circuit}. The magnetic field is swept for several different values of $V_0$, which has been scaled up to keep the voltage across the five parallel SMTJs similar to what it was for two~\footnote{The voltage supply is chosen as $V_0=V_\text{mean}\times[1+(2 N R_0)/(R_\text{P}+R_\text{AP})]$.  For the example shown in Fig.~\ref{fig:5chain}, $N=5$}. This fixes the simulation so that the coupling between each pair of SMTJs decreases as $1/N$, where $N$ is the number of SMTJs. 

\begin{figure}
	\def\svgwidth{\columnwidth}
    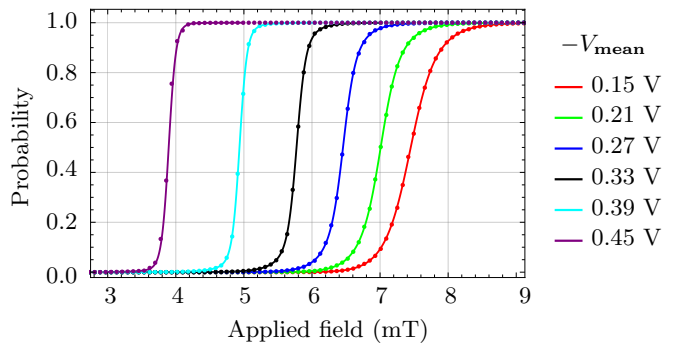
	\caption{\label{fig:5chain} Antiparallel state probability for each of five SMTJs electrically coupled in parallel as a function of applied magnetic field. The symbols give the probabilities computed with a Monte Carlo simulation (statistical error bars are smaller than the symbols) and the lines give the probabilities computed with a numerical extraction of the eigenvectors of the analytic Markov model. 
	}
\end{figure}

The two approaches for modeling the systems agree well; this is unsurprising because the Markov model should capture the the behavior of the simulations once they have reached steady state. For both approaches, the coupled SMTJs transition from all being in the parallel state to all being in the antiparallel states as the magnetic field is swept. As the applied voltage increases, the transitions shift to lower fields and occur over a narrower field range. The narrowing of the transition indicates increased coupling between the SMTJs. As the coupling increases, the SMTJs spend an ever increasing fraction of their time aligned with each other. 

Even though these simulations show that we can start building arrays of coupled SMTJs, we believe it is still the case that effectively using them will require more complicated coupling circuitry. So far, circuitry using SMTJs has involved schemes that fall into two broad groups. In the first, voltages are applied continuously across the SMTJs \cite{PhysRevX.7.031014,borders2019integer}. In the second, the SMTJ state is read and modified with voltage pulses, perhaps using a pre-charge sense amplifier \cite{PhysRevApplied.8.054045,mizrahi2018neural,daniels2020energy}. The former makes it easier for the states of different SMTJs to influence each other, but the latter can be more energy efficient~\cite{daniels2020energy}. For approaches with continuous voltages to reach the level of efficiency seen in pulsed approaches, the SMTJs must have a correlation time on the order of a nanosecond or less, and such devices are now under experimental development~\cite{safranski2020demonstration,hayakawa2021nanosecond}. The interacting $p$-bits demonstrated in Ref.~\cite{borders2019integer} output a digital bitstream and use analog input so that complicated circuitry is needed to couple the devices together. Ultimately, an energy-efficient approach will combine the advantages of both of the existing ones.

In this paper, we show that SMTJs can mutually couple through the electrical voltage stimuli caused by their stochastic electrical transitions.  This mutual interaction is established through a simple electrical connection, does not require complex circuitry, and is sufficient to modify the individual switching transitions of the two SMTJs. We believe that the coupling demonstrated with this compact approach is a useful starting point for building large assemblies of coupled SMTJs for novel cognitive computing schemes. The demonstrated ability of a simple Markov model to predict the coupled behavior of the SMTJs will allow predictive  modelling of SMTJs integrated with CMOS in a variety of approaches.

\begin{acknowledgments}
\label{Remerciements}
  We thank Robert McMichael, Paul Haney, and Daniel Gopman for enlightening discussions and their comments on the manuscript. AM and PT acknowledge support under the Cooperative Research Agreement Award No.~70NANB14H209, through the University of Maryland.
\end{acknowledgments}


\appendix
\label{appendix}




\section{N\'{e}el-Brown modelling of SMTJs}
\label{app:neelbrown}

\begin{figure*}
	\def\svgwidth{\textwidth}
    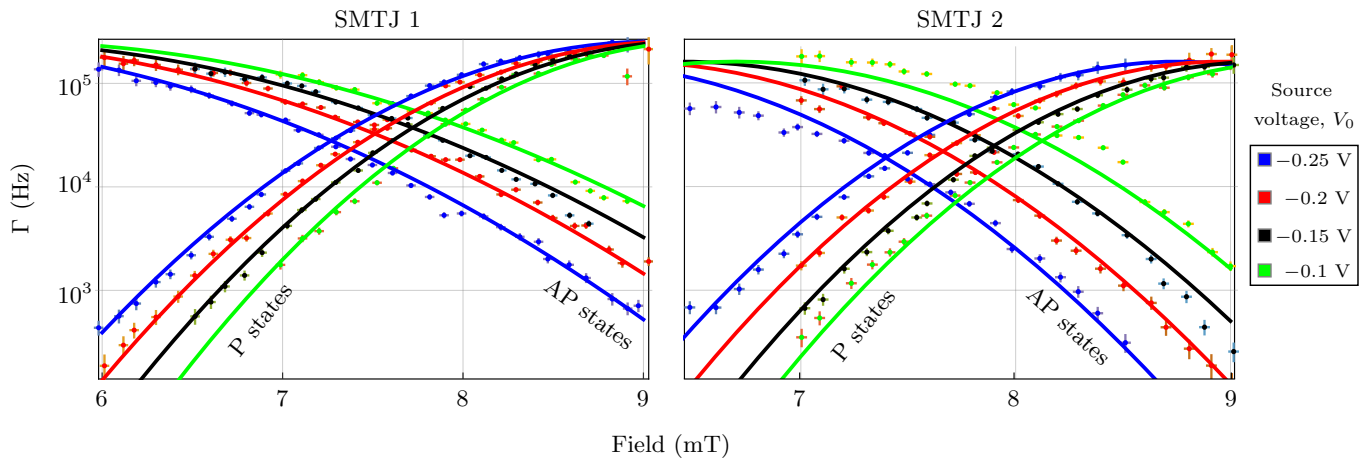
	\caption{\label{fig:fitDwellTime}Fits of the N\'eel-Brown model Eq.~\eqref{eq:neel-brown} to the uncoupled SMTJ transistion rates $\Gamma_{\pm}$ (e.g. Fig.~\ref{fig:single}) at four different source voltages.  Error bars indicate single standard deviation uncertainties due to the imprecision in setting the applied field value and the statistical uncertainty due to the limited number of transitions recorded for each point.}
\end{figure*}

To provide numerical input for modeling the coupled dynamics of the two SMTJs, we measure each SMTJ independently and uncoupled as in Fig.~\ref{fig:single}. From these measurements we extract the mean transition rates as a function of field for voltages $V_0\in\{-0.1,-0.15,-0.2,-0.25\}$~V. Fig.~\ref{fig:fitDwellTime} shows the measured results for each SMTJ.  These two SMTJs were selected from the wafer because their crossovers between parallel and antiparallel alignments occurred at similar fields and bias voltages; overlap of these crossing points maximizes the coupling.

To fit the measured transition rates, we use a N\'{e}el-Brown model that has been modified to include the effects of spin transfer torques~\cite{brown1963thermal,li2004thermally,li2008perpendicular,suh2008attempt,rippard2011thermal}.
The transition rates out of each state are given by
 \begin{align}
    \Gamma_{\pm } &=
 \Gamma_0 \exp\Bigg[-\beta\left(1 \pm \dfrac{V}{V_{c}}\right)\nonumber\\
 &\times \left(1 \pm \left[\dfrac{H-H^\pm_{0}}{ H^\pm_{k}}+AV+BV^2\right]\right)^2\Bigg], \label{eq:neel-brown} 
 \end{align}
where the plus sign is taken for antiparallel to parallel transitions and the minus sign for the opposite.
The voltages $V$ are extracted from the experimental time traces. Linear fits of SMTJ voltages $V$ as a function of source voltage $V_0$ allow us to extrapolate (and interpolate) beyond the range of collected data. All variables are fitting parameters and are given in Table~\ref{tab:table1} except for the applied field $H$ and the voltages $V$ which are the independent variables. Here $\beta=\Delta E/(kT)$ is nominally understood to be the ratio of the energy barrier $\Delta E$ of the SMTJ to the thermal energy ($T$ is the temperature, assumed to be fixed at $300$~K, and $k$ is Boltzmann's constant).  The prefactor, $\Gamma_0$, is sometimes taken to be $10^{9}$~Hz \cite{li2008perpendicular} and sometimes computed from the parameters of the Landau-Lifshitz-Gilbert equation \cite{suh2008attempt,kanai2021theory}, but here is treated as a fitting parameter, for reasons explained below. The critical switching voltage, $V_{c}$, incorporates the effect of the damping-like torque on the effective energy barrier and the parameters $A$ and $B$ include effects of the field-like torque and nonlinear effects. The offset magnetic field is $H_0^{\pm}$ and the anisotropy field is $H_{k}^{\pm}$.

\begin{table}
	\caption{\label{tab:table1}  N\'{e}el-Brown model parameters for SMTJ 1 and SMTJ 2 obtained from fits to the dwell times derived from the experimental voltage time traces for each device.  The set of values shown here is one of several roughly equivalent sets obtained from a high-dimensional fit with significantly nonorthogonal parameters.  These values should be considered useful for modeling, but should not be considered numerically definitive.   	}
	
	
	\begin{ruledtabular}
		\begin{tabular}{  l  r @{\hspace{-7ex}}  l r @{\hspace{-7ex}}  l}
    & \multicolumn{2}{c}{SMTJ 1}&  \multicolumn{2}{c}{SMTJ 2} \\ \cline{2-3} \cline{4-5}
		    $\Gamma_0$&\quad 0.254 & MHz  &  \quad 0.161  & MHz \\
		    $\beta$&4.34 &   &  6.55  & \\
		    $V_{c}$&$-0.55$ & V & $-2.7$   &V\\
		    $\mu_0H_{k}^{+}$& 2.11 & mT  & 2.31   &mT\\
		    $\mu_0H_{k}^{-}$& 4.15 & mT  & 2.78   &mT\\
		    $\mu_0H_{0}^{+}$& 7.32 & mT  &  7.27  &mT\\
		    $\mu_0H_{0}^{-}$& 9.76 & mT  &  9.86  &mT\\
		    $A$&$-0.50$ & V$^{-1}$  & $-1.7$   &V$^{-1}$\\
		    $B$&3.8 & V$^{-2}$ & 4.2   &V$^{-2}$\\ 
		\end{tabular}
	\end{ruledtabular}
\end{table}

We view this fitting scheme not as an attempt to extract physically correct values for various parameters, but as a physically plausible approach to obtaining parameter values for further modelling. The high dimensionality of the fit and nonorthogonality of the fit parameters lead to many degenerate solutions that give equally performant model predictions; the numerical values we report in Table~\ref{tab:table1} depend strongly on the initial values for the fitting routine, making uncertainties extracted from the nonlinear fit procedure meaningless. We find that using the same set of parameters for both orientations of the same SMTJ leads to poor fit performance, and that allowing different fields $H_0^{\pm}$ and $H_k^{\pm}$ for each state was the minimal extension needed to get good agreement with the data in Fig.~\ref{fig:fitDwellTime}. We conjecture that the different reversal parameters between the two states may follow from different magnetic configurations; the samples are low aspect ratio ellipses, and thus magnetic reversal may not follow single-domain switching dynamics, and the non-uniform fringing field from the pinned layer likely affects the two reversal processes differently. 

We also find that $\Gamma_0$ is needed as a fit parameter to reproduce the negative curvature of the experimental data, and regardless of initial value, $\Gamma_0$ converges to the fastest switching frequency present in the data: the values for $\Gamma_0$ in Table~\ref{tab:table1} correspond to the asymptotic maxima in Figs.~\ref{fig:fitDwellTime}. The discrepancy between this value of $\Gamma_0$ and the gigahertz value typically used in the literature means that our values of $\beta$, which nominally represent the energy barrier in units of $kT$, cannot be directly compared to values extracted with the faster assumed prefactor. To extract a meaningful energy barrier, we would need to measure temperature dependence of the SMTJ statistics. We cannot discount the possibility that some experimental artifact causes the observed saturation of the rates. However, the rate saturates at a value about two orders of magnitude smaller than the bandwidth of the measurement ($\approx 14$~MHz). Models~\cite{naaman2006poisson} of Poisson processes with finite measurement bandwidth suggest that the effect of such a bandwidth on the measured rates should be negligible. It may be that for near-circular samples such as these, the reversal mode may be very different than that for a macrospin and the prefactor consequently differing from that expected for such a system.

\section{Markov models for coupled SMTJs}
\label{app:markov}
 
In this Appendix, we review the theoretical model used to predict the coupled system's behavior from the uncoupled fits. The idea is to construct a Markov model based on the $2^N$ possible states of the system, where here the number of devices is $N=2$. In a continuous time Markov model, we need only consider single-device switching events, as the probability of two devices switching in the same infinitessimal $dt$ is $(dt)^2\rightarrow 0$. Therefore each transition rate in the Markov model can be understood from the switching of an isolated SMTJ, which is well-described by our N\'eel-Brown fits to the uncoupled SMTJ statistics, provided we account for the Markov-state-dependent voltage. The dependence of voltage on the states of the other SMTJs is the effective coupling mechanism, but the Markov process can be expressed purely using the uncoupled device fits. For our system, this model assumes that the state-dependent voltage arises instantaneously after a transition, which holds so long as the $RC$ times in the circuit are faster than the fastest SMTJ dwell time. With similar restrictions, the formalism we describe below can be applied to any system of coupled SMTJs so long as the coupled system experiences jump transitions between a discrete number of metastable states.
 
\begin{figure}
	\def\svgwidth{\columnwidth}
    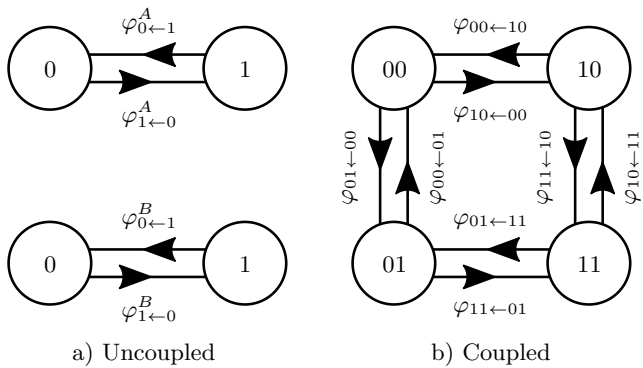
	\caption{\label{fig:Markovskya schema} a) Schematic of the model in the individual uncoupled case for the two different SMTJs; note the correspondence to Fig.~\ref{fig:circuit}. (1) or (0) corresponds to the parallel and antiparallel configuration of the SMTJ. Arrows represent jumping events characterized by a transition rate that forms the arrow label. b) Schematic of the model in the coupled case. Four different configurations are possible: (00), (01), (10), and (11). The first index corresponds to the state of SMTJ 1 while the second one is for the state of SMTJ2.
}
\end{figure}

Fig.~\ref{fig:Markovskya schema} illustrates schematically the model we consider to characterize the mutual coupling mechanism. In this approach, the $i^\text{th}$ uncoupled SMTJ can be described by two states and two transition rates $\varphi_{\pm,i}$ corresponding to the characteristic time elapsed before SMTJ $i$ escapes its $\pm$ state. In the case of two coupled SMTJs, there are four possible states: $(P,P)$, $(P,AP)$, $(AP,AP)$, and $(AP,P)$. In the rest of this section, we refer to these four states as 11, 10, 00, and 01 respectively to make the notation more compact. We consider transitions between these states as a continuous time Markov model so that eight nonzero transition rates $\{\varphi_j\}_{j=1}^8$ can be defined corresponding to the eight arrows in Figure.~\ref{fig:Markovskya schema}b. Note that due to the coupling, the rate for each SMTJ to transition out of a particular state depends on the state of the other SMTJ, so there are really 8 rates and not four as there would be in the absence of coupling. 

We can then define a $4\times 4$ transition rate matrix $M$ giving the transition rates between states, where $M_{ij} = \varphi_{j\rightarrow i}$. Eight of the sixteen matrix components are the nonzero transition rates; the diagonal elements are $M_{ii} =  - \sum_j \varphi_{i\rightarrow j}$, and the remaining four elements correspond to double transitions that have zero probability. The vector $\bm{P} = (P_{00},P_{01},P_{10},P_{11})$ indicating the conditional probabilities for the system to be found in each of the four states given some initial distribution $\bm{P}_0$ of the system then evolves according to the master equation $\dot{\bm{{P}}} = M\bm{P}$, which has the generic solution
\begin{equation}
 \bm{{P}}(t) = \exp(M t)\bm{P}_0.
\end{equation}
$M$ is diagonalizable, but as it has no other symmetries, its right and left eigenvectors differ in general. It is a singular matrix, so one of the eigenvalues is zero.  The other eigenvalues are all negative. We diagonalize $M$ as $M = V\Lambda V^{-1}$, with $\Lambda$ the diagonal matrix carrying the eigenvalues, $\lambda_j$. Then the evolution of $\bm{P}$ can be expressed simply as $\bm{{P}}(t) = V\exp(\Lambda t)V^{-1}\bm{{P}}_0$. The elements of $\bm{V}^{-1}$ and $\Lambda$ are all rational functions of the rates $\bm\varphi$ that can be computed analytically as solutions to the quartic characteristic equation of $M$ but these solutions are not notationally compact. 

The configuration $\bm{P}(t)$ can be expressed as a linear combination of the factors $e^{\lambda_j t}$ times the time-independent eigenvectors. Each element of $\bm{P}(t)$ corresponds to the probability density function of a hyperexponential distribution mixing the eigenvalues of $M$,
\begin{align}
  {P}_i(t) &= \sum_{j,k} e^{\lambda_j t} V_{ij} V^{-1}_{jk} {P}_{k}(0).\label{eq:master-solution}
\end{align}
In the experiment, we have no information about the state of the SMTJs before measurements begin. If we observe the system at time $t=0$, we can then declare the that ${P}_i(0) = \delta_{i\ell}$, indicating certainty that the system is in state $\ell$. Our knowledge about the likelihood of each state in the times $t>0$ then obeys the master equation above, and in particular $\bm{P}(t)$ will decay to $\bm{P}(\infty)$, the steady-state probability distribution of the system. This steady state distribution is the eigenvector of $M$ with eigenvalue zero. The other contributions in the sum over $k$ in Eq.~\eqref{eq:master-solution} are time-dependent and decay to zero on timescales $|\lambda_i|^{-1}$ because the remaining eigenvectors have negative eigenvalues. Only the $\bm{P}(\infty)$ contributes at long times after an observation. Since the coefficients under the sum of Eq.~\eqref{eq:master-solution} are readily computed numerically, we use the experimental fits from Table~\ref{tab:table1} together with our N\'eel-Brown model [Eq.~\eqref{eq:neel-brown}] to compute the expected form of $\bm{P}(t)$.

To analytically compute the cross-correlation $C(t)$ from the model, we  compute the Pearson correlation between the two states, one at time $t$ and one at time $0$, using the time-dependent probability distributions for each SMTJ extracted from $\bm{P}(t)$ by summing the probabilities over both possibilities of the state of the other SMTJ. The complicated rational forms for the solutions can be evaluated at the fit parameters extracted from the uncoupled N\'eel-Brown models even though the results offer little useful intuition. As a result of Eq.~\eqref{eq:master-solution}, the cross-correlation and each SMTJ's coupled autocorrelation are all superpositions of the same three exponentials arising from the eigenvalues of the transition rate matrix plus the time-independent steady state solution. The steady state solution is present in the same amount in all physical solutions because it is the only eigensolution in which the elements of the eigenvector do not sum to zero. This follows from the fact that at all times, the probabilities must sum to one and the steady state solution must always be the long time solution.

We consider two SMTJs with encoded time series data  $A(t)$ and $B(t)$ that correspond to the state of the devices at time $t$ with $A(t)=-1$ when the corresponding SMTJ is in the parallel state and $A(t)=1$ when it is in the antiparallel state.  The covariance of the two time series is 
\begin{align}
C_{A,B}(t) &=
\langle(A(t) - \langle A\rangle)(B(0)-\langle B\rangle)\rangle \\
&= \langle A(t)B(0)\rangle - \langle A\rangle\langle B\rangle   , 
\end{align}
where averages here are over Markov states. The terms in the covariance can be computed via the eigenspectrum of $M$ by using the solution to the master equation as a generator of conditional probabilities. For example, consider the calculation of $\langle A(t)B(0)\rangle$. This is by definition
\begin{align}
    \langle A(t)B(0)\rangle &= \sum_{\substack{i=\pm 1\\j = \pm 1}} i j P(A(t) = j , B(0) = i)\\
                            &= \sum_{\substack{i=\pm 1\\j = \pm 1}} i j P(A(t) = j  | B(0) = i) \nonumber\\
                            &\quad\quad\quad\quad\quad\quad\quad \times P(B(0) = i).
    \label{eq:crossterm}
\end{align}
The joint probability $P_{ji}=P(A(t) = j , B(0) = i)$ is the probability that both $P(A(t) = j$ and $B(0) = i$ are true and can be written in terms of the conditional probability $p_{ji}(t)=P(A(t) = j  | B(0) = i)$ that $A(t) = j$ given $B(0) = i$. The conditional probability is computed by assuming $B(0) = i$ as an initial condition for the master equation and examining the evolved distribution at time $t$. This is analogous to computing the two-point correlation function $\langle j | G(t,0) | i \rangle$ between two states in quantum mechanics, except that here our propagator is given by the solution to the master equation, rather than the Schr\"odinger equation. We have for the joint probability
\begin{align}
    P_{ji}(t) &= P(A(t) = j|B(0) = i)P(B(0) = i)\\
           &= \sum_\mu^{A(t) = j}\hat{\bm{e}}_{\mu}^T Ve^{\Lambda t}V^{-1}\bm{P}_\infty^{B = i} \label{eq:conditional}
\end{align}
where $\bm{P}_\infty^{B = i}$ is the steady state distribution marginalized over the assumption that SMTJ $B$ is in the $i$ state and $\hat e_{\mu}$ is the unit vector for the $\mu^\text{th}$ Markov state where the sum over $\mu$ is restricted to Markov states where SMTJ $A$ is in the $j$ state. As an example, if $i=1$ and $j=-1$, we might have $\bm{P}_\infty^{B=1} = (P_{00}, 0, P_{10}, 0)^T/(P_{00} + P_{10})$, and the two $\hat e_{\mu}$ of interest would be $(0, 0, 1, 0)$ and $(0,0,0,1)$. Using Eq.~\eqref{eq:conditional} as the conditional probability in Eq.~\eqref{eq:crossterm}, all that remains in the latter equation is to take the sum of the $ij=\pm 1$ weighted by the marginal probabilities $P(B_0 = i)$. 

Similar calculations based on using the master equation solution as a conditional probability can be used to compute any expectation values of interest. We used such expressions to generate the results shown in Fig.~\ref{fig:correlationFits}. From the perspective of Eq.~\eqref{eq:conditional}, we see that the curves in Fig.~\ref{fig:correlationFits} are all intrinsically sums of three exponentials (corresponding to the three nonzero eigenvalues in $\Lambda$). The apparent single-exponential behavior of the autocorrelations in that plot is only approximate, arising because the coefficient of one of the exponential terms dominates over the others. The inset to that figure shows that at long times, the time dependence of all three functions is dominated by the exponential with the longest decay time but with different prefactors.

The $t=0$ cross-correlation can be evaluated directly in terms of the stationary probabilities, without referring to any time-dependent evolution. By definition, the same-time correlation coefficient is
\begin{equation}
    C(0) = \frac{\langle A B \rangle-\langle A\rangle\langle B\rangle}{\sqrt{\sigma^2_A \sigma^2_B}} \label{eq:correlation-start}
\end{equation}
where $\sigma_{A,B}^2$ are the SMTJ-wise variances. Using our $\pm 1$ realizations of the SMTJ states, we have
\begin{align}
    \langle A \rangle &= P_{00} + P_{01} - P_{10} - P_{11}\\
    \langle B \rangle &= P_{00} - P_{01} + P_{10} - P_{11}\\
    \langle A B \rangle &= P_{00} - P_{01} - P_{10} + P_{11}
\end{align}
where $P_{i,j}$ is the probability of the joint state $(i,j)$ in the steady state limit. The numerator of Eq.~\eqref{eq:correlation-start}, which is just the covariance, then simplifies to $\text{cov}(A,B) = 4 \det \mathcal P$, where $\mathcal P$ is the matrix of Eq.~\eqref{eq:p-matrix} from the main text. To arrive at this form of the covariance it is convenient to multiply $\langle A B \rangle$ by $1 = \sum_{ij} P_{ij}$. As for the denominator, each of the variances can be written
\begin{align}
    \sigma^2_M &= \langle M^2\rangle - \langle M\rangle^2\\
    &= 1 - (P_0 - P_1)^2
\end{align}
where $P_1$ and $P_0$ are marginalized over the joint probabilities for the SMTJ of interest. The variances are then
\begin{align}
    \sigma^2_A &= 1 - [(P_{00}+P_{01}) - (P_{10} + P_{11})]^2\label{eq:variance-1}\\
    &= 4(P_{00} + P_{01})(P_{10} + P_{11})\label{eq:variance-2}\\
    \sigma^2_B &= 4(P_{00} + P_{10})(P_{01} + P_{11}).
\end{align}
In passing from Eq.~\eqref{eq:variance-1} to Eq.~\eqref{eq:variance-2}, it is useful to employ the identity $1 = (\sum_{ij} P_{ij})^2$. Taking these results together, the total correlation coefficient reduces to
\begin{equation}
    C(0) = \frac{P_{00}\times P_{11} - P_{01}\times P_{10}}{\sqrt{P_{0}^{(1)}\times P_{0}^{(2)}\times P_{1}^{(1)}\times P_{1}^{(2)}}}
    \label{eq:correlation-formula}
\end{equation}
where $P_j^{(k)}$ are the marginal probabilities for SMTJ $k$ (i.e. $P_{0}^{(1)} = P_{00} + P_{01}$).

\bibliography{references}

\end{document}